\documentclass[letterpaper,margin=1in]{optica-article}

\journal{osajournal} % for journals or Optica Open

\articletype{Research Article}

\usepackage{lineno}
\usepackage{graphicx}
\usepackage{cite}
\usepackage{comment}
\usepackage{braket}
\usepackage{dsfont}
\usepackage{xcolor}
\usepackage{subcaption}
\usepackage{caption}
\begin{document}

\title{Nonlocal Cancellation of Optical Rotations in Fructose Solutions}
%\title{Observation of Nonlocal Optical Rotation with Entangled Photons}
%\title{Entanglement-Enhanced Detection of Fructose in Solutions}

\author{Wen-Chia Lo,\authormark{1,2,*,$\dagger$} Chao-Yuan Wang,\authormark{1,2,*} Yu-Tung Tsai,\authormark{1,2} Sheng-Yao Huang,\authormark{1,2} Kang-Shih Liu,\authormark{1,2} Yun-Hsuan Shih,\authormark{1,2} Ching-Hua Tsai\authormark{1,2}, and Chih-Sung Chuu\authormark{$1,2,\ddagger$}}

\address{$^1$ Department of Physics, National Tsing Hua University, Hsinchu 30013, Taiwan\\
$^2$ National Center for Excellence in Quantum Information Science and Engineering, National Tsing Hua University, Hsinchu 30013, Taiwan
}

\email{\authormark{$\dagger$}g225149@gmail.com}
\email{\authormark{$\ddagger$}cschuu@phys.nthu.edu.tw}
\email{\authormark{*} These authors contributed equally to this work.}%% email address is required; see note below about the corresponding author designation

% use {asbstract*} to suppress the copyright line. Copyright information will be added in production

\begin{abstract*} 
Entanglement, one of the most representative phenomena in quantum mechanics, has been widely used for fundamental studies and modern quantum technologies. In this paper, we report the observation of nonlocal cancellation and addition of optical rotations with polarization-entangled photons in fructose solutions. The entanglement also enables probing optical activities at a distance by joint measurements on the entangled photons. The good agreement between the experimental results and theoretical predictions demonstrates the potential for extending these measurements to other chiral molecules, with a sensitivity that improves as the number of entangled photons increases.
%The good agreement between the experiment and the theory shows the potential for generalizing the measurements to other chiral molecules, with a sensitivity that will increase with the number of entangled photons. 

%This work explores the application of quantum entangled photon pairs in non-local measurements and their capability to analyze optically active materials. By leveraging post-selected cavity SPDC, Bell states ($\ket{\psi_{+}}$ and $\ket{\psi_{-}}$) are generated with concurrence values of 0.875 , demonstrating strong entanglement properties. The non-locality of the photon pairs is confirmed through CHSH inequality measurements, consistently exceeding the quantum bound. These quantum probes are used to measure polarization rotations induced by fructose solutions of varying concentrations, achieving superior efficiency by extracting average and differential properties with a single non-local measurement. This approach highlights the potential of quantum probes in surpassing classical methods for advanced sensing applications.

\end{abstract*}

%%%%%%%%%%%%%%%%%%%%%%%%%%  body  %%%%%%%%%%%%%%%%%%%%%%%%%%
\section{Introduction}
Entanglement of photons has been widely used in various quantum technologies, ranging from quantum communication \cite{Zapatero2023} and quantum computing \cite{OBrien2007} to quantum sensing \cite{Degen2017}. In quantum key distribution (QKD) \cite{EkertPRL1991,BennettPRL1992}, entanglement allows remote users to share secret keys with the security guaranteed by quantum mechanics. In linear optical quantum computing \cite{Knill2001}, entanglement plays a key role in increasing the success rates of quantum gates through quantum teleportation \cite{BennettPRL1993} and in reducing the overhead of resources with cluster states \cite{Raussendorf2001}. In quantum sensing, entanglement is instead exploited to improve the sensitivity of measurements beyond the standard quantum limit. For example, super-resolving phase measurements are demonstrated with multiphoton entangled states \cite{MitchellN2004}. %The precision in polarimetry can also be enhanced by using entangled photons \cite{Pedram2024}.

Polarization-entangled photons have also been used to probe the optical activity and optical rotatory dispersion exhibited by a solution of chiral molecules \cite{NoraScienceAdvances2016}, and to improve precision and sensitivity in polarimetry \cite{Pedram2024}. Optical rotation, or circular birefringence, is a manifestation of the optical activity of a chiral medium through the polarization plane rotation of the incident linearly polarized photons. In the case of solutions of chiral molecules with one or more chiral centers, a molecule with a specific stereochemical configuration rotates the plane of polarization in one direction, while its mirror image (its enantiomer) rotates the plane by the same amount in the opposite direction. If the two
enantiomers are present in different quantities, the total optical rotation will be determined by the
enantiomeric excess between the two forms, which makes optical rotation a useful optical phenomenon capable of recognizing and distinguishing the two forms, as the enantiomers otherwise have the same
chemical and physical properties.  

In this paper, we demonstrate the fascinating non-classical features of quantum sensing on optical rotations in fructose solutions by reporting the first observation of nonlocal optical rotation cancellation and addition of polarization-entangled photons, an analog to nonlocal dispersion cancellation~\cite{FransonPRA1992,Valencia2002,BaekOE2009} or nonlocal modulation \cite{Harris2008,Sensarn2009,Belthangady2009,Wu2019} of time-energy-entangled photons. We also demonstrate the possibility of probing the optical activities of two chiral media, as simulated by wave plates for greater control and deeper analysis, at a distance (in two buildings) and simultaneously by joint measurements on the transmitted entangled photons. Classically, determining the optical rotations in two solutions would require two separate measurements. The excellent agreement between our observation and theory, as well as the optical activities measured by the entangled photons and a laser beam, manifests the potential for applying the techniques to different chiral molecules or to a higher number of entangled photons.

\section{Nonlocal cancellation and addition of optical rotations}

To measure the optical rotation, linearly polarized laser beams are commonly used to compare the orientations of the polarization before and after passing through the sample.

The same circular birefringence of chiral molecules is also experienced by single photons. The optical rotation of a linearly polarized photon can be modeled as a unitary operator $\hat{U}(\theta)=\text{e}^{-i\hat{\sigma}_{y}\theta}$ \cite{Goldberg2021}, with $\hat{\sigma}_j$ being the Pauli $j$ matrix, which corresponds to a rotation about the $y$-axis on a Bloch sphere. The rotation angle $\theta$ of the polarization can be acquired by measuring both the expectation values of $\hat{\sigma}_{z}$ and $\hat{\sigma}_{x}$ locally, 
with the resolution limited by the shot noise if single photons are exploited. 

In the following, we show that \textit{nonlocal cancellation} and \textit{addition} of optical rotations may be observed if entangled photons are used to probe optical rotations at a distance.

Consider a pair of photons in the maximally entangled states (Bell states), $\ket{\psi_{\pm}}=\frac{1}{\sqrt{2}}(\ket{\text{H}}_{\rm A}\ket{\text{V}}_{\rm B}\pm\ket{\text{V}}_{\rm A}\ket{\text{H}}_{\rm B})$, where $\ket{\text{H}}_k$ ($\ket{\text{V}}_k$) represents the horizontally (vertically) polarized state of the photon entering the solution of chiral molecules at the location $k=$ A or B.
After each photon passes through one sample corresponding to a unitary operator $\hat{U}_k$ with optical rotation $\theta_k$, the entangled state becomes
\begin{equation}
\begin{aligned}
\ket{\psi'_{\pm}}=\ &\hat{U}_{\rm A}(\theta_{\text{A}})\otimes\hat{U}_{\rm B}(\theta_{\text{B}})\ket{\psi_{\pm}}\\
=\ &\frac{1}{\sqrt{2}}[\mp\sin(\theta_{\text{A}}\pm\theta_{\text{B}})\ket{\text{H}}_{\rm A}\ket{\text{H}}_{\rm B}+\cos(\theta_{\text{A}}\pm\theta_{\text{B}})\ket{\text{H}}_{\rm A}\ket{\text{V}}_{\rm B}\\ &\pm\cos(\theta_{\text{A}}\pm\theta_{\text{B}})\ket{\text{V}}_{\rm A}\ket{\text{H}}_{\rm B}+\sin(\theta_{\text{A}}\pm\theta_{\text{B}})\ket{\text{V}}_{\rm A}\ket{\text{V}}_{\rm B}]\\
=\ &\hat{U}_{\rm A}(\theta_{\text{A}}\pm\theta_{\text{B}})\otimes\hat{U}_{\rm B}(0)\ket{\psi_{\pm}}\\
\equiv\ &\hat{U}_{\rm A}(\theta_{\pm})\otimes\hat{U}_{\rm B}(0)\ket{\psi_{\pm}}\\
\end{aligned}
\label{nonlocal}
\end{equation}
as if one of the entangled photons passed through both solutions at once and obtained a \textit{nonlocal} optical rotation of $\pm\theta_{\rm B}$, of which the sign can be controlled by the entangled state, in addition to a \textit{local} optical rotation of $\theta_{\rm A}$. Consequently, when $\theta_{\text{B}}=\theta_{\text{A}}$, entangled photons in the state $\ket{\psi_{-}}$ experience no optical rotation ($\theta_{\text{-}}=0$) as if both solutions were optically inactive, exhibiting what we term the nonlocal cancellation of optical rotations. Another interesting feature is that, if $\theta_{\text{A}}>\theta_{\text{B}}$ ($\theta_{\text{A}}<\theta_{\text{B}}$), the photon with nonlocal optical rotation would experience levorotation $\theta_{\text{-}}>0$ (dextrorotation $\theta_{\text{-}}<0$) even though both solutions are dextrorotatory $\theta_{\text{A,B}}<0$ (levorotatory $\theta_{\text{A,B}}>0$). For the entangled photons in the state $\ket{\psi_{+}}$, the photon with nonlocal optical rotation would experience an optical rotation $\theta_{+}=\theta_{\text{A}}+\theta_{\text{B}}$ that will be magnified by a factor of 2 if $\theta_{\text{A}}=\theta_{\text{B}}$. 

The effect of nonlocal cancellation or addition of optical rotations with entangled photons can be observed through the joint measurements %\textcolor{red}{through the quantity $M_{zz}^{\pm}$ and $M_{xz}^{\pm}$ defined as:}
%through the joint measurements in the $\sigma_{z}\otimes\sigma_{z}$ or $\sigma_{x}\otimes\sigma_{z}$ basis,%
\begin{equation}
\begin{aligned}
M_{zz}^{\pm}&\equiv\langle \psi'_{\pm} | \hat{\sigma}^{\rm A}_{z}\otimes\hat{\sigma}^{\rm B}_{z} | \psi'_{\pm} \rangle
%=\text{tr}(\sigma_{z}\otimes\sigma_{z}\rho_{\psi'_{\pm}})
=-\cos(2\theta_{\pm}),\\
M_{xz}^{\pm}=\pm M_{zx}^{\pm}&\equiv\langle \psi'_{\pm} | \hat{\sigma}^{\rm A}_{x}\otimes\hat{\sigma}^{\rm B}_{z} | \psi'_{\pm} \rangle
%=\text{tr}(\sigma_{x}\otimes\sigma_{z}\rho_{\psi'_{\pm}})
=-\sin(2\theta_{\pm}),
\end{aligned}
\label{joint}
\end{equation}
where $\hat{\sigma}^k_j$ is the Pauli matrix operated at location $k$, with the extrema independent of local optical rotations.

In comparison, the extrema of the joint measurements using non-entangled photons in the separable state $\ket{\phi_{sep}}=\ket{\text{H}}_{\rm A}\ket{\text{V}}_{\rm B}$,
%$\rho_{\text{HV}}=\ket{\text{HV}}\bra{\text{HV}}$ 
\begin{equation}
\begin{aligned}
M^{sep}_{zz}\equiv&\langle \phi_{sep} | \hat{\sigma}^{\rm A}_{z}\otimes\hat{\sigma}^{\rm B}_{z} | \phi_{sep} \rangle
%&=\text{tr}(\sigma_{z}\otimes\sigma_{z}\rho_{\text{HV}})
=\cos(2\theta_{\text{A}})\cos(2\theta_{\text{B}}),\\
M^{sep}_{xz}\equiv&\langle \phi_{sep} | \hat{\sigma}^{\rm A}_{x}\otimes\hat{\sigma}^{\rm B}_{z} | \phi_{sep} \rangle
%&=\text{tr}(\sigma_{x}\otimes\sigma_{z}\rho_{\text{HV}})
=-\sin(2\theta_{\text{A}})\cos(2\theta_{\text{B}}),
\end{aligned}
\label{classical}
\end{equation}
would depend on local optical rotations. 

The nonlocal optical rotation also enables the detection of optical activity over a distance. If $\theta_{\text{A}}$ and $\theta_{\text{B}}$ are both unknown within $\pm\pi/4$, their values can be extracted through joint measurements by
\begin{equation}
\begin{aligned}
\theta_{\rm A, B}=-\frac{i}{4}\text{ln}[(-M_{zz}^{+}-iM_{xz}^{+})(-M_{zz}^{-}-i\epsilon_{\rm A, B} M_{xz}^{-})],
%\theta_{\text{B}}&=\frac{-i}{\textcolor{red}{4}}\text{ln}[(-M_{zz}^{+}-iM_{xz}^{+})(-M_{zz}^{-}+iM_{xz}^{-})].
\end{aligned}
\label{phases}
\end{equation}
where $\epsilon_{\rm A} = - \epsilon_{\rm B} =1$. An unknown $\theta_{\rm A}$ within $\pm\pi$ can be found by varying $\theta_{\text{B}}$ nonlocally to maximize $|M_{zz}^{-}|$ or minimize $|M_{xz}^{-}|$. 

To see how entanglement may enhance the sensitivity of nonlocal measurement of optical activity, let us consider an $N$-photon entangled state $(\ket{\text{R}}^{\otimes N} - \ket{\text{L}}^{\otimes N})/\sqrt{2}$, where $\ket{\rm R} = (\ket{\rm H}-i\ket{\rm V})/\sqrt{2}$ and $\ket{\rm L} = (\ket{\rm H}+i\ket{\rm V})/\sqrt{2}$. If each photon passes through a solution of the same optical rotation $\theta$, the output state will be $\ket{\psi}=(e^{iN\theta}\ket{\text{R}}^{\otimes N} - e^{-iN\theta}\ket{\text{L}}^{\otimes N})/\sqrt{2}$. The quantum Fisher information \cite{Sidhu2020} corresponding to the nonlocal optical rotation measurement is then $F_Q(N)=4[\langle \frac{\partial}{\partial \theta} \psi | \frac{\partial}{\partial \theta} \psi \rangle - |\langle \psi | \frac{\partial}{\partial \theta} \psi \rangle |^2 ]=4N^2$, which gives a lower bound of standard variance $1/F_Q(N)$ scaling as $O(1/N^2)$ while the standard variance using $N$ non-entangled photons would scale as $O(1/N)$.

\section{Experiments}
\subsection{Nonlocal optical rotation in fructose solutions}

\begin{figure}[h!]%H为当前位置，!htb为忽略美学标准，htbp为浮动图形
\centering %图片居中
\includegraphics[width=1\textwidth]{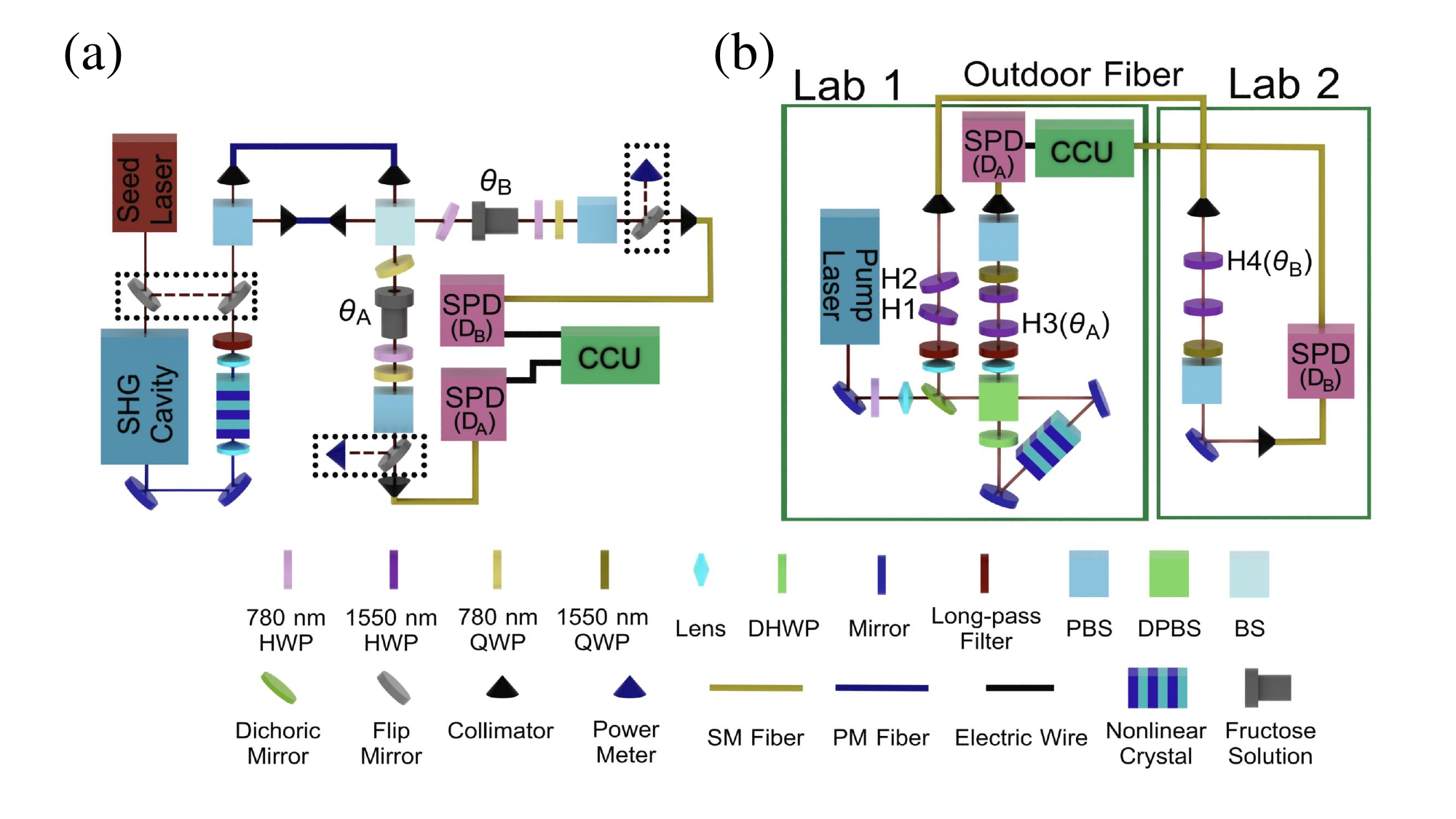}
\caption{
Experimental setups for observing nonlocal cancellation and addition of optical rotations with entangled photons. (a) In the experiment with fructose solutions, 795 nm polarization-entangled photons are generated using cavity-enhanced SPDC and a beam splitter (BS) via post selection \cite{Wu2017,Wu2019,Cheng2020}. 
A quarter-wave plate (QWP) and half-wave plate (HWP) are then used to adjust the phase between $\ket{\text{H}}_{\rm A}\ket{\text{V}}_{\rm B}$ and $\ket{\text{V}}_{\rm A}\ket{\text{H}}_{\rm B}$. 
(b) In the experiment across two buildings, polarization-entangled photons at 1535 and 1560 nm are generated with a thermally stabilized PPLN crystal in a Sagnac interferometer \cite{KimPRA2006}. 
A 773.8 nm picosecond laser is exploited as the pump laser.  
The polarization analyzers in both setups consist of a HWP, a QWP and a polarization beam splitter (PBS) before the photons are collected into the single-photon detectors (SPDs). Coincident events are analyzed and recorded by time digitizers (CCUs).
}
\label{Fig.setup 2} %用于文内引用的标签
\end{figure}

Figure~\ref{Fig.setup 2}(a) illustrates the experimental setup for observing nonlocal cancellation and addition of optical rotations in fructose solutions. A 397.5 nm pump laser, frequency-doubled from a 795 nm seed laser, is shone onto a periodically poled potassium titanyl phosphate (PPKTP) crystal to emit bright biphotons at 795 nm with a bandwidth of 4.5 MHz \cite{Wu2017,Wu2019,Cheng2020}. The biphotons are then sent into a 50:50 beam splitter (BS) to create a polarization-entangled state with each photon exiting at different ports \cite{ShihPRL1988}. The polarization-entangled photons are subsequently sent into two fructose solutions separated by 62 cm. Each solution, sealed in a 7 cm long glass container, is prepared by dissolving fructose (Sigma Aldrich, CAS number 57-48-7) in
purified water with a molarity (the number of moles of fructose per liter of solution) of $0 \sim 4.236$ M. 

The transmission of entangled photons through solutions is 75\% and is independent of the polarizations or 
molarities used in our experiment. To implement the joint measurements in Eq.~(\ref{joint}), we use polarization analyzers, composed of a half-wave plate (HWP), a quarter-wave plate (QWP), and a polarization beam splitter (PBS), to select the appropriate bases.

\begin{figure}[t]
    \centering
    \includegraphics[width=1\textwidth]{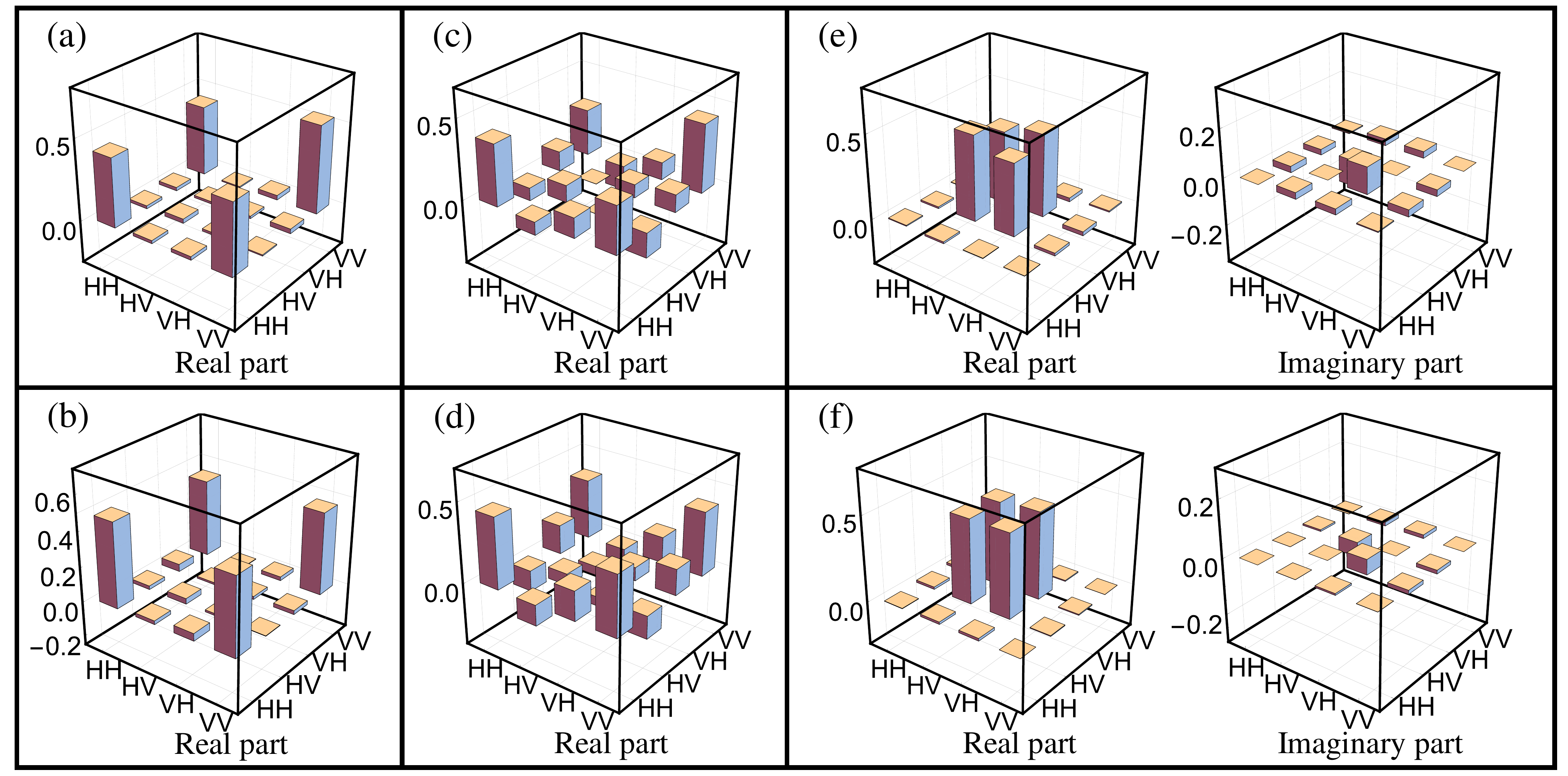}
    \caption{Reconstructed density matrices of the polarization-entangled photons with one photon passing through (a) air (where the state is prepared in $\ket{\phi_{+}}=\frac{1}{\sqrt{2}}(\ket{\rm H}_A \ket{\rm H}_B+\ket{\rm V}_A \ket{\rm V}_B)$), (b) purified water, and (c) fructose solution (of which the molarity corresponds to an optical rotation of $\theta_A = 20.08^{\circ}$). (d) The predicted density matrix of the entangled photons, in which the presence of fructose is modeled by $\hat{U}\left(\theta_{\rm A}\right)=e^{-i{\hat{\sigma}}_y\theta_{\rm A}}$, has a cosine similarity of 0.94 compared to the density matrix in (c). The reconstructed density matrices of Bell states $\ket{\psi_{+}}$ used in the experiments with fructose solutions (measured in front of the solutions) and across two buildings (measured at the input of the 300-m-long fiber) are shown in (e) and (f), respectively. 
     }
    \label{HV+VH tomo}
\end{figure}

Figures~\ref{HV+VH tomo}(a) to \ref{HV+VH tomo}(c) show the reconstructed density matrices of polarization-entangled photons with one photon passing through air, purified water, and fructose solution, respectively. The reconstruction of density matrices uses quantum state tomography (QST) \cite{JamesPRA2001}. A set of coincidence counts is first measured in 16 tomographic analysis bases $\ket{\alpha}_{\rm A}\ket{\beta}_{\rm B}$ with $\{\alpha, \beta \} = \{\rm H, \gamma \}$, $\{\rm V, \gamma \}$, $\{\rm R, \gamma \}$, and $\{\rm D, \delta \}$, where $\gamma = \{\rm H, V, D, L \}$, $\delta = \{\rm H, V, D, R \}$, and $\ket{\rm D} = (\ket{\rm H}+\ket{\rm V})/\sqrt{2}$. The maximum likelihood estimation is then used to numerically find the density matrix most likely to have produced the measured data set. As shown in Fig.~\ref{HV+VH tomo}(d), the presence of fructose [compared to Fig.~\ref{HV+VH tomo}(c)] is well described by the operation of $\hat{U}(\theta)=\text{e}^{-i\hat{\sigma}_{y}\theta}$.

Figure~\ref{HV+VH tomo}(e) shows the reconstructed density matrix of the photon pairs in $\ket{\psi_{+}}$, with a fidelity of $F=0.917$ and a concurrence \cite{WoottersPRL1998} of $C=0.874$. The deviation from an ideal Bell state results mainly from the asymmetric wavepacket of the entangled photons and the uncorrelated photon pairs. 

We also characterize the optical activity of fructose solutions using the linearly polarized 795 nm seed laser, which has the same wavelength as the entangled photons, with a power meter. 

\begin{figure}[t]
    \centering

    % 第一排
    \hspace{-0.03\textwidth} % 向左靠一點
    \begin{subfigure}{0.45\textwidth}
        \begin{minipage}[t]{\linewidth}
            \text{(a)}\\
            \includegraphics[width=\linewidth]{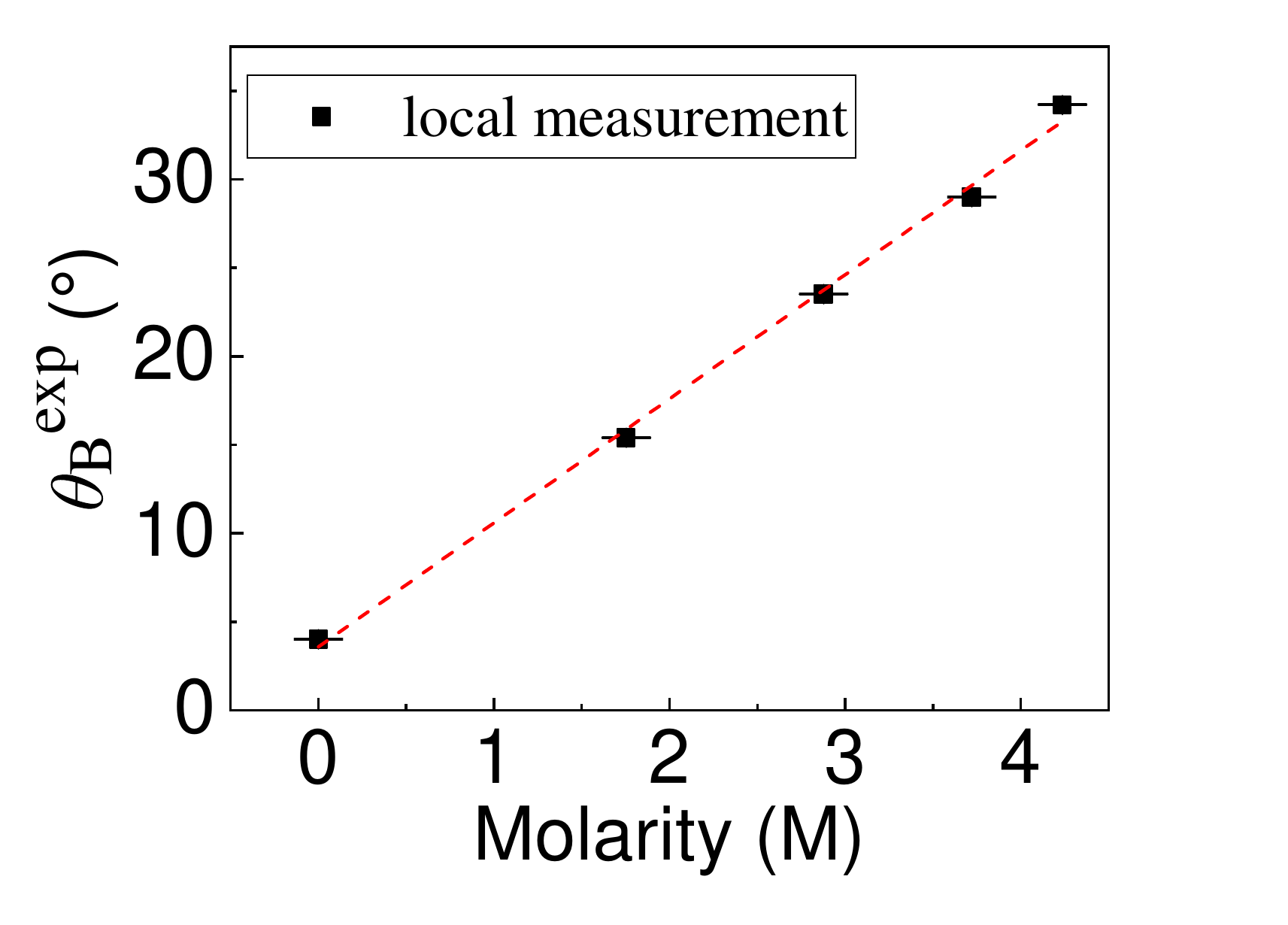}
        \end{minipage}
    \end{subfigure}
    %\hfill
    \hspace{-0.03\textwidth} % 向左靠一點
    \begin{subfigure}{0.45\textwidth}
        \begin{minipage}[t]{\linewidth}
            \text{(b)}\\
            \includegraphics[width=\linewidth]{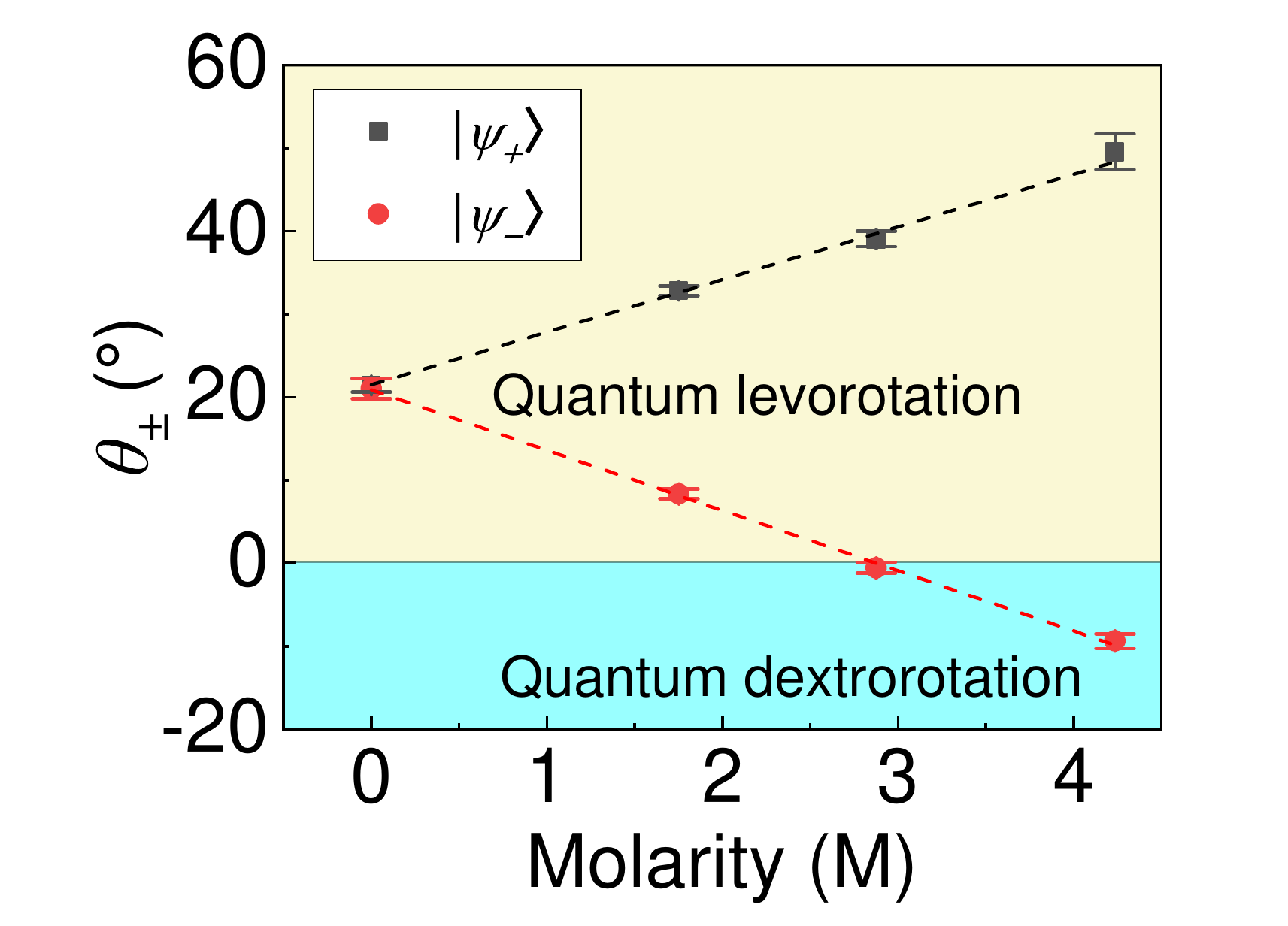}
        \end{minipage}
    \end{subfigure}

    \caption{
    (a) Optical rotation of a linearly polarized laser beam measured for different  %mass fractions%
    molarities of fructose solution. The standard deviation is $\pm 0.04^{\rm o}$. (b) Measurement of $\theta_{\pm}$ with entangled photons prepared in $\ket{\psi_{\pm}}$. The shaded yellow and blue regions indicate quantum levorotation ($\theta_{\pm} > 0$) and quantum dextrorotation ($\theta_{\pm} < 0$), respectively. 
    All data are available in the Supplementary Materials.
    }
    
   \label{Non-local fructose measurement graph}
\end{figure}

As shown in Fig.~\ref{Non-local fructose measurement graph}(a), the fructose solution induces a levorotatory rotation $\theta_{\rm B}^{\rm exp}(c) = \theta_{\rm B}(c) + \theta_{\rm B}^{\rm PBS} = (7.01 ^{\circ}/{\rm M}) c + 4.09^{\rm o}$ proportional to the % $mass fraction $w$ %
molarity $c$ of fructose. The offset $\theta_{\rm B}^{\rm PBS}$ (or $\theta_{\rm A}^{\rm PBS} = -4.75^{\rm o}$ in the $\theta_{\rm A}$ measurement) in the absence of fructose is due to the misalignment between the axes of the PBS and the wave plates in the polarization analyzers.

To observe nonlocal cancellation and addition of optical rotations, we keep the %mass fraction%
molarity of fructose solution A constant at %51.8\% %
2.877 M (or $\theta_{\rm A}=20.08^{\circ}$) while varying the %mass fraction %
molarity of solution B. Figure~\ref{Non-local fructose measurement graph}(b) shows the observed $\theta_{\pm} = \theta_{\pm}^{\rm exp} - \theta_{\rm A}^{\rm PBS} \mp \theta_{\rm B}^{\rm PBS} - \theta^{\rm HWP}$ for various %mass fractions% 
molarities of solution B, where the offsets ($\theta_{\rm A}^{\rm PBS}$ and $\theta_{\rm B}^{\rm PBS}$) due to the misalignment of the axes in the polarization analyzers and the rotation ($\theta^{\rm HWP} = 5.47^{\rm o}$) induced by the wave plate for exchanging the $\ket{\psi_{\pm}}$ states (only with $\theta_-^{\rm exp}$) are corrected from the measured rotations $\theta_{\pm}^{\rm exp}$.

For the entangled photons in the state $\ket{\psi_{+}}$ (black squares), $\theta_{+}$ is observed to increase with %mass fraction%
molarity and is always larger than $\theta_{\rm A}$, a phenomenon we term nonlocal addition of optical rotation. Moreover, $\theta_{+}$ always has the same sign (levorotation) as $\theta_{\rm A}$ or $\theta_{\rm B}$. In contrast, for the entangled photons in the state $\ket{\psi_-}$ (red circles), $\theta_-$ is levorotatory ($\theta_->0$) as single photons for the %mass fraction below 51.8\% %
molarity below 2.877 M (or $\theta_{\rm B}<\theta_{\rm A}$) but becomes dextrorotatory ($\theta_-<0$) for the %mass fractions above 51.8\% %
molarity above 2.877 M (or $\theta_{\rm B}>\theta_{\rm A}$). When both solutions have the same %mass fraction of 51.8\% % 
molarity of 2.877 M (or $\theta_{\rm B}=\theta_{\rm A}$), we observe nonlocal cancellation of optical rotations ($\theta_- = 0$). These observations agree well with the theory.

\subsection{Nonlocal optical rotation over a long distance}

The observation of nonlocal cancellation and addition of optical rotations is independent of distance. To demonstrate such a feature, we place the sources of optical rotation separately in two buildings on the campus of National Tsing Hua University in Taiwan. One of the entangled photons generated in the General II Building is sent to the Physics Building via a 300-m-long single-mode fiber. To minimize the transmission loss (0.188 dB/km at 1550 nm), polarization-entangled photons are generated with center wavelengths of 1535 and 1560 nm using a Sagnac interferometer \cite{KimPRA2006
%,FedrizziOE2007,Liu2022
} [Fig.\ref{Fig.setup 2}(b)]. The phase between the $\ket{\rm H}_{\rm A}\ket{\rm V}_{\rm B}$ and $\ket{\rm V}_{\rm A}\ket{\rm H}_{\rm B}$ states at the output of the fiber is also constantly monitored and stabilized using a polarization controller and a tilted wave plate. 
A superconducting nanowire single-photon detector (SNSPD) and an avalanche single-photon detector are employed with a time digitizer to analyze the incoming signals. A pair of tilted HWPs (H1 and H2) is installed before the detectors to compensate for the relative phase and alter the Bell state between $\ket{\psi_{+}}$ and $\ket{\psi_{-}}$.

The polarization entanglement of the emitted photon pairs is first examined by the QST and the CHSH inequality. Figure \ref{HV+VH tomo}(f) shows the reconstructed density matrix of the photon pairs in $\ket{\psi_{+}}$, with a fidelity of $F=0.984$ and a concurrence of $C=0.968$. The CHSH inequality \cite{BellPPF1964,ClauserPRL1969} is given by $S = \vert E(a, b) - E(a, b')\vert + \vert E(a', b)\vert + \vert E(a', b')\vert\leq2$, where $E(x,y)=P_{00}(x,y)-P_{01}(x,y)-P_{10}(x,y)+P_{11}(x,y)$ is the correlation function and $P_{mn}(x,y)$ is the probability that detectors D$_
{\rm A}$ and D$_
{\rm B}$ both click when the polarization analyzers at location A and B are set at the angles of $x=\{a, a'\}$ and $y=\{b, b'\}$ (for $m=0$ and $n=0$) or $x=\{a+90^{\rm o}, a'+90^{\rm o}\}$ and $y=\{b+90^{\rm o}, b'+90^{\rm o}\}$ (for $m=1$ and $n=1$), respectively. The inequality only holds if the states satisfy the local-hidden-variable theory \cite{EinsteinPR1935}. A maximally entangled state violates the inequality with $S=2\sqrt{2}$ for properly chosen angles $a, a',b$, and $b'$. In our experiment, $\{a,a'\}=\{0^\circ,45^\circ\}$ and $\{b,b'\}=\{22.5^\circ,67.5^\circ\}$. The measured $S$ is $2.818\pm0.0094$, which violates the inequality by 87 standard deviations.

\begin{figure}[tp] 
\centering 
\includegraphics[width=0.9\textwidth]{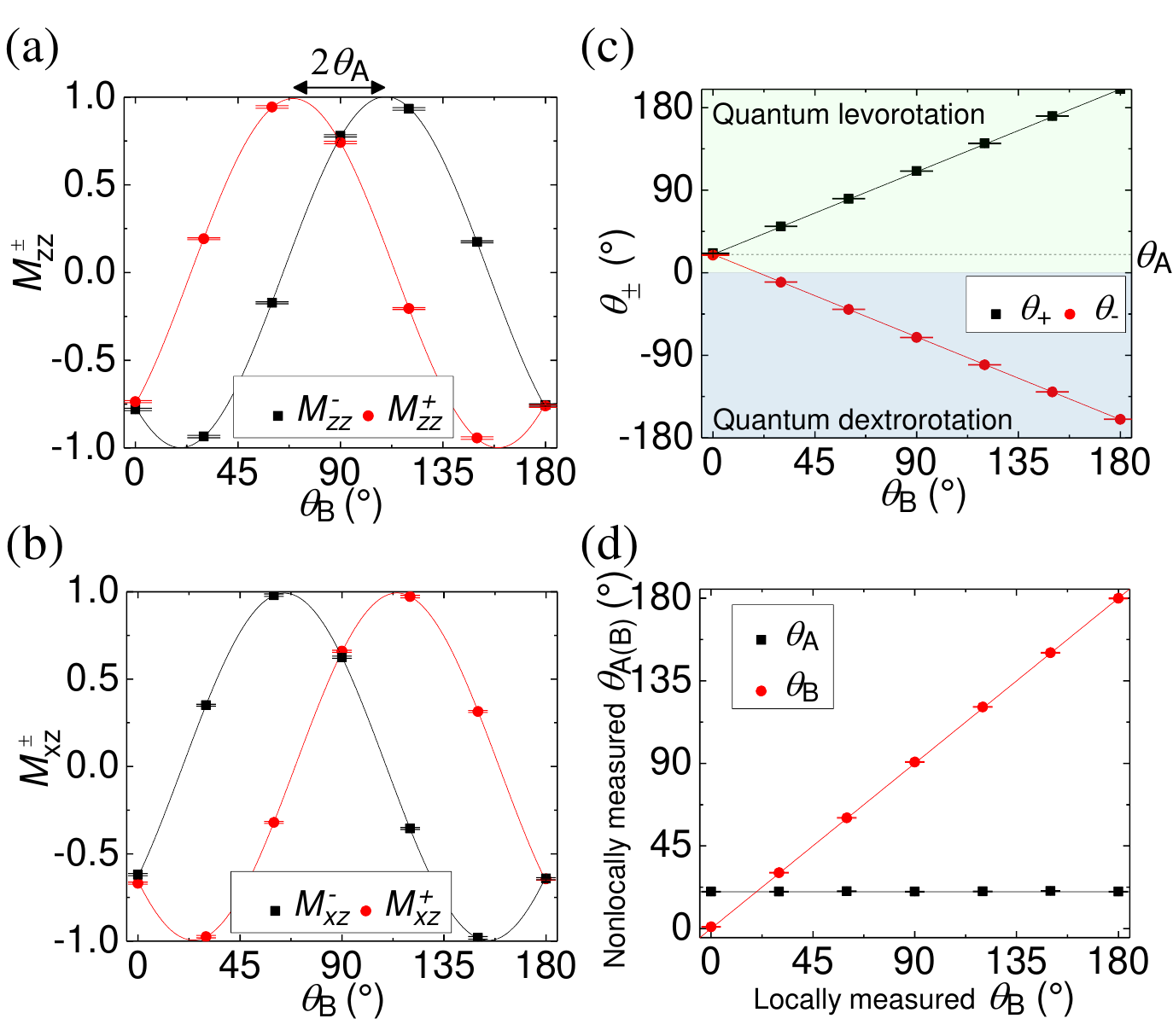} 
\caption{Joint measurements (a) $M^{\pm}_{zz}$ and (b) $M^{\pm}_{xz}$ for different $\theta_{\text{B}}$ and $\theta_{\text{A}}=20^\circ$. The curves are the fitting using trigonometric functions. The high visibility implies the nonlocal effect of entangled photons. (c) Observation of nonlocal cancellation ($\theta_-=0$) and nonlocal addition of optical rotations. The standard deviation is $\pm 0.3^{\rm o}$. (d) Nonlocal measurements of $\theta_{\text{A}}$ (square) and $\theta_{\text{B}}$ (circles) compared to the experimentally prepared $\theta_{\text{A}}=20^{\circ}$ (black line) and $\theta_{\text{B}}$ (red line). The standard deviation is $\pm 0.21^{\rm o}$. All data are available in the Supplementary Materials.}
\label{Expectation value curve} 
\end{figure}

To achieve comprehensive control and analysis of nonlocal optical rotations, we experimentally simulate the optical activities of chiral molecules using motorized zero-order HWPs (H3 and H4). We vary the optical rotation in one building by rotating the angle $\theta_{4}$ of H4 and keep the angle of H3 in another building at $\theta_3=\theta_{\text{A}}/2$. Figures~\ref{Expectation value curve}(a) and \ref{Expectation value curve}(b) show the obtained $M_{zz}^{\pm}$ and $M_{xz}^{\pm}$ as functions of $\theta_{\text{B}}=2\theta_{4}$, with entangled photons prepared in $\ket{\psi_{+}}$ (red dots) and $\ket{\psi_{-}}$ (black dots), respectively, and $\theta_{\text{A}}=20^\circ$. The observation agrees well with the theoretical predictions of Eqs.~(\ref{nonlocal}) and (\ref{joint}). The obtained $M_{zz}^{\pm}$ and $M_{xz}^{\pm}$ are cosine and sine functions of $2\theta_\text{B}$ (red and black curves), with the curves of $M_{zz}^{\pm}$ (or $M_{xz}^{\pm}$) separated from each other by $2\theta_{\rm A}$. The upper and lower bounds of $M_{zz}^{\pm}$ or $M_{xz}^{\pm}$ at $\pm1$, a manifestation of entanglement, are also in contrast to separable states, which would be reduced by a factor of $\cos(2\theta_{\text{A}})$ or $\sin(2\theta_{A})$ and without separation of $2\theta_{\text{A}}$ [Eq.~(\ref{classical})]. Figure ~\ref{Expectation value curve}(c) shows the measured $\theta_{\pm}$ as a function of $\theta_{\text{B}}$. Again, nonlocal cancellation and addition of optical rotations are observed with the entangled photons in $\ket{\psi_{-}}$ (red circles) and $\ket{\psi_{+}}$ (black squares), respectively. Comparison to the theoretical values of $\theta_{\text{A}}\pm\theta_{\text{B}}$ (black and red lines) gives a coefficient of determination $R^2$ of 0.99995 and 0.99996 for $\theta_{+}$ and $\theta_{-}$, respectively, which confirms the theory. Finally, with the obtained $M_{zz}^{\pm}$ and $M_{xz}^{\pm}$, we can also detect the values of $\theta_{\rm A}$ and $\theta_{\rm B}$ over a distance. The results are given in Fig.~\ref{Expectation value curve}(d), in which $\theta_{\rm B}$ is varied and $\theta_{\rm A}$ is kept constant at $20^{\circ}$. Compared with the experimentally prepared $\theta_{\rm B}$ (red line) and $\theta_{\rm A}=20^{\circ}$ (black line), the measured $\theta_{\rm B}$ (circles) and $\theta_{\rm A}=20.22^{\circ}$ (squares) are found to have $R^2=$ 0.9995 and 0.99998, respectively, showing good agreement.

\section{Conclusion}
We have demonstrated for the first time nonlocal cancellation and addition of optical rotations with polarization-entangled photons in fructose solutions, and the possibility of probing the optical activities of two chiral media at a distance and simultaneously by joint measurements on the transmitted entangled photons. The good agreement between the experiment and the theory, as well as the measurements using entangled photons and a laser beam, demonstrates the potential of applying our techniques to other chiral molecules (such as glucose, blood sugar, proteins, lipids, or solids) or with a higher number of entangled photons. We note that narrowband entangled photons \cite{Wu2017,Cheng2020} are used in our experiment to reduce dispersion-induced optical rotation in fructose solutions. This dispersion effect may be further suppressed using entangled photons with a narrower bandwidth \cite{Chinnarasu2020,Lin2025}. We also note that our technique can be miniaturized for compact and precise chiral sensing with a low sample volume using lab-on-a-chip technology such as microfluidics \cite{Whitesides2006}.

\section*{Funding}
National Science and Technology Council, Taiwan (113-2119-M-007-012, 114-2119-M-007-012); Ministry of Education (Top Research Centers in Taiwan Key Fields Program). 

\section*{Acknowledgments}
The authors would like to thank D.-S. Chuu and C.-Y. Yang for the insightful discussion and endless support in every aspect.

\section*{Disclosures}
The authors declare no conflicts of interest.

\section*{Data availability}
Data underlying Figs.~\ref{Non-local fructose measurement graph} and \ref{Expectation value curve} are included in Supplement 1. Data underlying other results presented in this paper are not publicly available at this time but may be obtained from the authors upon reasonable request.

\section*{Supplemental document} See Supplement 1 for supporting content.

\medskip

\bibliography{Nonlocal_measurement_ref_20241210}

\end{document}